\documentstyle[12pt]{article}
\global\arraycolsep=2pt
\newcommand{\be}{\begin{eqnarray}}
\newcommand{\ee}{\end{eqnarray}}
\newcommand{\la}{\langle}
\newcommand{\ra}{\rangle}
\begin{document}
\begin{titlepage}
\begin{flushright}
hep-ph/9604407\\
BROWN-HET-1035\\
April 1996
\end{flushright}
\vspace{0.3cm}
\begin{center}
\Large\bf
Can Asymptotic Series Resolve the Problems of Inflation?
\end{center}

\vspace {0.3cm}

\begin{center}    {\bf Robert H. Brandenberger}\footnote
{e-mail address:rhb@het.brown.edu}
\end{center}
\begin{center}
{\it Physics Department, Brown University, Providence, RI 02912, USA}
\end{center}
\begin{center}
{\it and}
\end{center}
\begin{center}
{\bf Ariel R. Zhitnitsky}\footnote{
e-mail address:arz@physics.ubc.ca }
\end{center}
\begin{center}
{\it Physics Department, University of British Columbia,
6224 Agricultural  Road, Vancouver, BC V6T 1Z1, Canada}
\end{center}
\begin{abstract}
We discuss  a cosmological scenario in which inflation is driven by a
potential which is motivated by an effective Lagrangian approach to gravity.
We exploit the recent arguments \cite{ARZ}
that an effective Lagrangian $L_{eff}$
which,  by definition,
contains operators of arbitrary dimensionality
is in general not a convergent but rather an asymptotic series
with factorially growing coefficients.  This
behavior of the effective Lagrangian might be responsible
for the resolution of the cosmological constant problem.
We argue that the same behavior
of the potential gives  a natural realization of the inflationary
scenario.
\end{abstract}
\end{titlepage}
\vskip 0.3cm
\noindent

\setcounter{page}{1}
\section{ Introduction}

Inflation \cite{Guth} \cite{Linde} is an attractive paradigm for early Universe cosmology.
Not only does it provide a mechanism to render the observable part of the Universe approximately spatially flat and homogeneous, but it also gives rise to a mechanism for the origin of inhomogeneities on galactic and super-galactic scales \cite{Flucts}. In spite of these successes, however, a convincing realization of inflation in the context of fundamental physics is still missing. In most models of inflation, the existence of a new fundamental scalar field $\phi$ must be postulated, and, in addition, it
's potential $V(\phi)$ is tightly constrained by cosmological considerations. The presence of such scalar fields also makes the cosmological constant problem worse since the value of the minimum of $V(\phi)$ is not set by any particle physics considerations.

In this Letter we suggest an inflationary scenario in which the inflaton, the scalar field $\phi$ giving rise to inflation, is a scalar condensate $\Phi$ arising in an effective field theory description of gravity. We determine its potential, making use of an asymptotic series analysis \cite{ARZ} of the effective field theory and show that some of the problems of inflation are resolved in a natural manner. A second crucial ingredient of our mechanism
is a dynamical infrared cutoff.

Note that both gravitational effects and condensates have previously been invoked to generate inflation. In fact, the first model of inflation \cite{Starobinsky} made use of an $R^2$ correction term in the gravitational action. More recently, a realization \cite{Slava} of inflation based on a
special class of higher derivative gravity models implementing the Limiting Curvature Hypothesis \cite{Markov} was proposed. Condensates were also considered in the past \cite{Ball}, \cite{Parker} as a mechanism to drive inflation.

The first crucial new aspect of our analysis is the asymptotic series analysis of the effective field theory \cite{ARZ} in order to discover the potential for the condensate. It turns out that the potential is very different from the potentials typically
used to generate inflation. Since expectation values of condensates of
massless fields such as the graviton are infrared divergent, it is required to introduce an infrared cutoff. In the context of cosmology, this infrared cutoff
is not arbitrary. The corresponding length scale is constrained by
physical considerations (see e.g. \cite{Linde2}). Taking into account this
time dependent infrared cutoff is the second key new aspect of this work.

Let us recall some general arguments \cite{ARZ}
regarding the effective description
of gravity at the Planck scale.
We assume that at some stage in the very early Universe the gravity field as well as
other relevant fields (e.g. scalar fields)
take on nonzero vacuum expectation values (VEV),
which we shall call condensates  \footnote{The condensates will disappear at
late times, as discussed later in the text.}.
In gravity this condensation process is likely to occur at the Planck scale,
and it can be viewed
in analogy with the  phenomenon of
gluon condensation in QCD which takes place at a scale of $1 GeV$ \cite{Gluon}.
We denote any relevant condensate by $\la  \Phi \ra$.  It might stand for a  scalar field like the ones
cosmologists often introduce to describe inflation (the inflaton), the dilaton or a gravitational field  $\la R \ra $ itself (the restriction to scalar condensates is just for notational convenience).
The natural scale for such a condensate
is, of course, the  Planck scale. For the higher dimensional
operators $Q^n$ which depend on $\Phi$  we assume
that there is a  factorization rule which
allows us to estimate the higher order condensates in the following way:
$\la Q^n \ra \sim \la \Phi^n \ra \sim \la \Phi \ra^n$.
This assumption is not crucial for our purposes.

As discussed in \cite{ARZ}, the fact that the perturbative expansion of
arbitrary Greens functions for the fundamental field theory is asymptotic \cite{Lipatov}, with factorially growing coefficients, leads to the conclusion that the effective Lagrangian
constructed from  the high dimensional operators
is  also an asymptotic series
in $\Phi$. Hence, given the above assumptions, the vacuum expectation value
of the
energy of the condensate field can be expressed as an asymptotic series in
$\la \Phi \ra $:
\be
\label{1}
\la H_{eff}\ra = \sum_{n=0}^{n=\infty}n!(-1)^na_n \la\Phi\ra^n,
\ee
with coefficients $a_n \sim 1$. We can use the Borel representation formula to evaluate the
asymptotic series for $\la H_{eff} \ra$ \footnote{The assumption that the
series is Borel summable is, as we argued in
\cite{ARZ}, not crucial for our analysis.}. Defining a function $f(s)$
through
\be
\label{2}
a_n {n!} = \int_0^{\infty} f(s)s^{-n-2}\exp(-\frac{1}{s})ds
\ee
(for all values of n) we obtain
\be
\label{3}
\la H_{eff}\ra = \int_{0}^{\infty}\frac{f(s)ds}{s(s + \la\Phi\ra)}
\exp(-\frac{1}{s}) ,
\ee
If the coefficients $a_n$ are not changing too fast,
then the function $f(s)$ will be
slowly varying, which we will assume to be the case in the following.

An example of an effective Lagrangian of the above form arises in QCD
(see   \cite{ARZ}). The
effective Lagrangian for collective degrees of freedom (e.g. colorless
mesons) is, after integrating
out high frequency quark and gluon degrees of freedom, an asymptotic series
when expanded in powers of the collective field.

It is known
that the higher order quantum gravity corrections to
physical observables in a de Sitter phase of an expanding Universe
are in general infrared divergent.
In particular,
this divergence is observed in the
vacuum correlator $ \la\phi^2\ra $
(see the recent papers \cite{Woodard}-\cite{Dolgov}
on this subject and references to previous publications therein).
This divergence
leads to corrections to expectation values which have power-law and not exponential dependence on time $t$.
This means that nonperturbative dynamics
should be taken into account.
However, such dynamics is as yet unknown.
Therefore we follow \cite{ARZ} and introduce
a phenomenological cutoff parameter $\epsilon(t)$
into the vacuum expectation value (VEV)
\be
\label{4}
\la\Phi\ra \rightarrow\frac{\la\Phi\ra}{\epsilon(t)
} ,
\ee
in order to account for the new physics
responsible for removing the infrared divergences mentioned above.
In the context of cosmology, we naturally expect this physics and
therefore $\epsilon$ to depend on time.

The introduction of the new field $\epsilon(t)$ is an absolutely
crucial ingredient of our approach.  This is, as  mentioned earlier,
an attempt to take
into account infrared physics which is not fully understood.
It is possible that the function $\epsilon(t)$
is related to the zero modes.
These modes should be treated separately, and their
dynamics may be described by a new classical
effective field $\epsilon (t)$.
At this point,
the dynamics of this new field is not known,
and the field magnitude is also uncertain. However
(and this is our main point) we do know that one and the same
function (one and the same physics) is responsible for explaining
both phenomena under discussion: inflation and the cosmological constant
problem. It may be the case (in the spirit of stochastic chaotic inflation \cite{Linde}) that the initial conditions in different
universes give rise to different behaviors of the function $\epsilon(t)$,
and that in some universe some of the
conditions to be discussed in the next section are not satisfied.
In this case inflation would continue forever, resulting
in a completely empty and cold Universe with nonzero cosmological
constant.
For our Universe the function $\epsilon (t)$ has some specific properties,
in particular it is close to zero today. Exactly such a function,
satisfying some constraints,
we keep in mind in what follows\footnote{We are grateful
to Andrei Linde for very useful discussions regarding this subject.}.

Making use of (\ref{4}), the integral which describes the vacuum energy
becomes
\be
\label{5}
\la H_{eff}\ra \sim
\int_{0}^{\infty}\exp(-\frac{1}{s})
\frac{ f(s)ds}{s(s  + \frac{\la\Phi\ra}{\epsilon(t)}
)}\sim\epsilon\rightarrow 0
\ee
and goes to zero if $\epsilon$ tends to zero.
Thus, the cosmological constant problem associated with the field driving inflation is reduced
to the problem of analyzing the function $\epsilon(t)$.
As we demonstrated in
\cite{ARZ}, the effect (\ref{5}) does not depend crucially on
our assumptions concerning both the factorization properties of
the condensates
$\la \Phi^n \ra \sim \la \Phi \ra^n$ and the exact factorial dependence of the coefficients.
Both of these effects presumably can be modeled (as we have done) by introducing
in formula (\ref{3}) a smooth function $f(s)$
whose moments  exactly reproduce the given
$n-$dependence of the coefficients.
If this function is mild enough, it will not destroy the relation
(\ref{5}), but might change some numerical coefficients.
Besides that, a condensate might have both a singular piece
proportional to
$\frac{\la\Phi\ra}{\epsilon(t)}$ and a regular part, i.e. be
proportional to
$\frac{\la\Phi\ra}{\epsilon(t)}+ {\rm const}$.
As can be seen, this does not invalidate the relation (\ref{5}).

A few remarks are in order.
The vanishing of the vacuum energy is a consequence  of the
asymptotic nature of the effective Lagrangian and of the infrared properties
of the VEVs
(which are hidden in the dynamics of the function $\epsilon(t)$) and it
does not require any fine tuning of parameters.
The vanishing of the vacuum energy (\ref{5}) can be interpreted
(after inflation, when all relevant  condensates
presumably go to zero) as the vanishing of the cosmological constant,
which is
the only relevant operator in the effective Lagrangian
(all other terms  are marginal or irrelevant operators).

We close this section with the following remark: the
strong infrared dependence of the vacuum condensate
$ \la\Phi\ra $ is not a unique property of quantum gravitational effects in de Sitter space.
Two-dimensional QCD with a large number of colors
also exhibits a strong
infrared dependence of a form similar to what we have postulated in (\ref{5}). In particular,
the so-called mixed vacuum condensates can be exactly
calculated in this theory in the chiral
limit ($m_q\rightarrow 0$). They exhibit the following
dependence on the infrared parameter $m_q$ \cite{Chibisov}:
\be
\label{6}
\frac{1}{2^n} \la  \bar{q}(ig\epsilon_{\lambda\sigma}
G_{\lambda\sigma}  \gamma_5)^nq \ra
=(-\frac{g^2\la\bar{q}q\ra}{2m_q})^n\la\bar{q}q\ra ,
\ee
where $q$ is a quark field and $G_{\lambda\sigma}$ is a gluon
field of $QCD_2(N=\infty)$. The chiral
condensate $\la\bar{q}q\ra$ in this theory can be calculated
exactly \cite{ARZ1}.  It does not vanish
without contradicting Coleman's theorem.
The very important feature of this formula: it diverges
in the chiral limit $m_q\rightarrow0$, where the parameter
$m_q$ plays the role of the infrared regulator of the theory.
Now , if we consider the
asymptotic series  constructed from these condensates, then
\be
\label{7}
\sum_{n=0}^{n=\infty} (-)^nn! a_n
\la  \bar{q}(ig\epsilon_{\lambda\sigma}
G_{\lambda\sigma}  \gamma_5)^nq \ra
\sim\int_{0}^{\infty}\exp(-\frac{1}{s})
\frac{dsf(s)}{s(s+\frac{1}{m_q})}\sim m_q\rightarrow 0,
\ee
(with a function $f(s)$ related to the coefficients $a_n$ as before by (\ref{2}))
we would get a result of zero  for this series,
in spite of the fact that each term
on the left hand side  diverges
in the chiral limit and irrespective of the precise behavior of
the coefficients $a_n$ in the formula (\ref{7})!
It should be clear that this result is a direct consequence of the
asymptotic nature of the series.
Although this example is only a toy model,
it gives us  a hint of what might happen in real Nature.

\section{The Condensate as an Inflaton}

In this section we will discuss
the possibility that inflation can be realized
within the same framework of an asymptotic series analysis of the
Effective Lagrangian which we briefly reviewed in the Introduction.
We shall demonstrate that inflation can be obtained
in a very natural way, driven by the same condensate
$\la\Phi\ra$ which we discussed earlier.

The general ideology is the  following.
One could expect that at a very early epoch a phase transition
similar to the confinement-deconfinement
transition in QCD occurs.
 In the case of QCD  we know that a gluon
condensate is formed in the infrared region and disappears
at higher temperatures. Such a behavior of a condensate is a
welcome property for the inflationary scenario
(see Ref.\cite{Ball} where the idea of the inflaton being a condensate
  was also considered, but from a different point of view).
If an analogous  phenomenon could occur for
$ \la\Phi\ra $ it would provide the chance to describe inflation
without introducing any new fields and specific potentials for them.
Rather,   inflation would be described
in terms of the nontrivial dynamics of the gravitational field itself.
A nonzero condensate can lead to inflation
because it provides a nonzero potential. The vanishing of the condensate
with time (as a consequence of the dynamics of
all relevant fields) will automatically lead
to a termination of the epoch of inflation.

How one could describe such a scenario in a more quantitative way?
The honest answer is that we do no know, because such a description
is inevitably based on quantum gravity at the Planck scale.
However, one may try to use some phenomenological input to the problem.
Thus, we suggest to consider the VEV of the energy (\ref{5})
as an {\it effective} potential of the theory
with the condensate $\la\Phi\ra$.
As we discussed earlier, this condensate $\la\Phi\ra$ is not necessarily
the condensate of a scalar field (which may not exist as a fundamental
field in the theory), but rather an effective description of the
strength of the interactions of all relevant fields.
Therefore, with the motivation given above, we propose
to interpret the potential (\ref{5})
as an effective potential $V_{eff} (\Phi)$  which is the essential element
of an inflationary cosmology:
\be
\label{8}
V_{eff} (\Phi) = {\tilde c}
\int_{0}^{\infty}\exp(-\frac{1}{s})
\frac{f(s)ds}{s(s m_{pl} + \frac{ \Phi }{\epsilon(t)})},
\ee
with the constant $\tilde c$ setting the energy scale at which inflation
takes place. We normalize $\epsilon(t)$ such that initially $\epsilon (0)
 = 1$.
More precisely, we interpret this potential
as an effective interaction which is an inherent element
of the classical equation of motion describing
the evolution of the condensate \cite{Linde}, \cite{Robert}:
\be
\label{9}
\ddot{\Phi}+3H\dot{\Phi}+V_{eff}^{\prime}(\Phi)=0,
\ee
where $\prime$ denotes the derivative with respect to $\Phi$.

The main question addressed in this Letter is whether the
condensate $\la\Phi\ra$ (denoted by $\Phi$ from now on) can act as the inflaton.
We will show that the standard conditions for
the inflationary potential
(like the ``slow rolling" condition) can easily be satisfied
for a potential of the form (\ref{8}). Before we proceed,
let us note that the potential (\ref{8})
is very different from one what people usually consider.
In particular, it  is very flat,
has no minimum  for finite $\Phi$, and for large $\Phi$ goes like
$1/\Phi$. However, these properties
 should not be considered
as a sign that the theory is ill defined. As we discussed before,
$ V_{eff} (\Phi) $ is not a fundamental potential
describing a fundamental field. It must be considered as an effective one.
Thus, there is no requirement for it to have a global minimum.
This potential, by definition, is not designed
to quantize a theory starting from the well-defined vacuum state.
Rather, by construction, it should be thought as giving the effective
strength of all relevant interactions.

We now turn to the analysis of inflation driven by a condensate with potential $V(\Phi)$ given by (\ref{8}). In order to obtain a viable inflationary cosmology, several requirements must be satisfied. In order to obtain inflation at all, the ``slow rolling" conditions must be obeyed. After inflation, there must be a period in which the energy of the inflaton is efficiently transferred to matter fields (this is the ``reheating" requirement). A stringent constraint on all inflationary theories is that they
not produce an amplitude of density perturbations in excess of the observational limits (the ``fluctuation problem"). Finally, it must be verified that there are well justified initial conditions which lead to inflation.

In our scenario, the condensate $\Phi$ forms at some very early time with a value $\Phi \simeq 0$. Spatial fluctuations in the value of $\Phi$ will not dominate the energy density. Hence, provided that the ``slow rolling" conditions are satisfied, inflation will set in, and the spatial gradients will be
exponentially suppressed. Thus, soon after the formation of the condensate, its equation of motion is well described by (\ref{9}).

A sufficient condition for inflation is that the ``slow rolling" conditions
\be
\label{10}
V^{\prime} m_{pl} \ll \sqrt{48 \pi} V
\ee
and
\be
\label{11}
V^{\prime \prime} \ll 24 \pi V m_{pl}^{-2}
\ee
are satisfied. The first condition expresses the requirement that the potential energy of $\Phi$ dominate over the kinetic energy (and one thus gets an inflationary equation of state), the second criterion ensures that the $\ddot \Phi$ term in the equation of motion (\ref{9}) can be neglected (and one thus gets a period of inflation of a sufficient number of Hubble expansion times in order to solve the horizon and flatness problems \cite{Guth}).

Let us first neglect the time dependence of $\epsilon$.
It is easy to show that for $\Phi \ll \epsilon$ the slow rolling conditions are satisfied. Since $f(s)$ is slowly varying and the exponential factor in (\ref{8}) cuts off the s-integral at $s \simeq 1$, the potential $V(\Phi)$ can be approximated as follows:
\be
\label{12}
V(\Phi) \simeq {\tilde c} \int_1^{\infty}ds s^{-2} f(s) \sim {\tilde c} f(1).
\ee
Analogous approximations for $V^{\prime}(\Phi)$ and $V^{\prime \prime}(\Phi)$ show that the slow rolling conditions (\ref{10}) and (\ref{11}) are satisfied independent of the value of ${\tilde c}$.

The slow rolling conditions are satisfied also for large values of $\Phi$ ($\Phi > \epsilon$). In this case, the potential $V(\Phi)$ can be approximated as
\be
\label{13}
V(\Phi) \simeq {\tilde c} \int_1^{s_\Phi}ds {{f(s)} \over {s \Phi / \epsilon}}
\simeq {\tilde c} {{f(s_\Phi)} \over {\Phi / \epsilon}} \rm{ln} (s_\Phi),
\ee
where $s_\Phi$ is the value of $t$ for which
\be
\label{14}
s_\Phi = {{\Phi} \over {\epsilon}}.
\ee
The quantities $V^{\prime}(\Phi)$ and $V^{\prime \prime}(\Phi)$ can be estimated in a similar fashion, with the result that they are suppressed compared to (\ref{13}) by $1/\Phi$ and $1/(2\Phi^2)$, respectively. Thus, the inequalities (\ref{10}) and (\ref{11}) are again obeyed independent of the value of
${\tilde c}$.

{}From the above discussion is appears that the ``slow rolling" conditions are satisfied for all values of $\Phi$. If this were true, the resulting cosmological scenario would be unacceptable since the scale factor would keep on inflating (albeit with a decreasing expansion rate), resulting in a completely empty and cold Universe today.

Post-inflationary reheating is a necessary requirement for a successful inflationary cosmology. In the usual models of inflation, the ``slow rolling" approximation breaks down after a finite duration of inflation. Afterwards, the inflaton starts oscillating about its ground state. The inflaton oscillations induce parametric resonance \cite{Parametric1}, \cite{Parametric2} in the fields which couple to $\Phi$, leading to rapid energy transfer from the inflaton to matter fields.

How can a graceful exit from inflation and reheating be realized in our scenario? The time dependence of the infrared cutoff $\epsilon$
may  play a crucial role. At the beginning of inflation, we expect the effective infrared cutoff to be given by the Hubble radius (see \cite{Linde2} for the use of this infrared cutoff in a different context), i.e.
\be
\label{15}
\epsilon(t) \propto  H (t).
\ee
The value of $\epsilon$ will decrease as $\Phi$ increases and the Hubble expansion rate $H(t)$ decreases. Well into the inflationary epoch, our system must be analyzed as a two field problem, the fields being $\Phi(t)$ and $\epsilon(t)$. Our system then
becomes similar to
Linde's two field inflation model \cite{Linde1}, although the physical origin of the second field is quite different.
We believe that there is no classical equation for $\epsilon(t)$ (beyond what is assumed in (\ref{15})). Its behavior is completely determined by the quantum nature of gravity.
There is one more essential new element in comparison
with Linde's paper \cite{Linde1}: our effective potential
(\ref{8}) is very different from
a standard potential for the scalar fields $\sim \Phi_1^2\phi_2^2$
which are finite polynomials with well defined minima.
As we mentioned earlier, our potential is not designed
to quantize a theory starting from a well-defined vacuum state;
rather it should be thought as an effective potential
motivated by the form of the asymptotic series(\ref{1})
of all relevant fields. Of course, one could start from the very beginning
from the  two field model given by potential (\ref{8})
without any motivation concerning the origin of the $\Phi(t)$ and
$\epsilon(t)$ fields. The technical analysis in this case
would proceed along the lines of \cite{Linde1}. However, we believe
that the motivation for the introduction of our fields is
a very important issue even if we do not completely understand
the dynamics of those fields.

There are indications
\cite{Woodard} that the time dependence of expectation values of massless
fields in an initial de Sitter state has corrections which depend on a power $p$ of time.
Such a behavior is realized during the time when quantum
corrections are in fact corrections, i.e. they are small.
When they get large, we can no longer use a perturbative series. Rather,
in order to describe $\epsilon(t)$
we must take into account all terms in the corresponding expansion at once.
Clearly, we do not know how to do this at the moment.
Instead, we assume a specific behavior for the function $\epsilon (t)$
with the following general feature: it is arbitrary function which
decreases with time and
vanishes at some finite time $t_c$.
To be specific, we shall assume
\be
\label{16}
\epsilon(t) = \epsilon(0) [ 1 - a (H t)^p ],
\ee
where $t = 0$ is the initial time, $p$ is an integer greater than 1, and $a$
is a constant much smaller than 1.
For the two field system, the condition (\ref{10}) required to obtain an
inflationary equation of state becomes
\be
\label{17}
{\dot \epsilon}^2 + {\dot \Phi}^2 < 2 V(\Phi).
\ee
Initially, the kinetic term for $\epsilon$ is subdominant. However, after
a number of expansion times greater than $a^{-1}$, this kinetic term begins
to dominate. Eventually it exceeds the potential energy. At this point, inflation ends.

We have thus shown that under certain assumptions about the time evolution
of $\epsilon$, our model yields a graceful exit from inflation after a
number of e-foldings which can be made large by decreasing the value of $a$.
The specific manner in which the Universe reheats remains to be analyzed.
We note that the
vanishing of $\epsilon(t)$ at some time $t_c$
is an important feature of the function $\epsilon(t)$. This is
because such a behavior enables us to obtain $\dot {\epsilon}^2 >> H^4$ while
$\epsilon$ itself is close to zero. After inflation ends,
the Universe approaches the
time $t_c$ when $\epsilon(t_c)=0$. In the vicinity
of this point our series (\ref{8})
is completely reconstructed: condensates  disappear and an
effective potential in the form (\ref{8})
does not make any sense. The existence of the function $\epsilon$ itself
is also terminated.  Right after this point we enter the stage
when gravity is a weekly coupled theory and when the cosmological
constant (vacuum energy) is almost zero\footnote{The magnitude of
the vacuum energy is proportional
to $\epsilon$. Thus, the precise     magnitude of the
cosmological constant depends on
the dynamics of the field $\epsilon$. During the
period of inflation the inertia of this field is very large
because the kinetic term  is very small for ``slow rolling'' conditions
to be satisfied. The same inertia pushes the field
$\epsilon$ further and further to zero. It is not clear to us
whether $\epsilon$ will actually intersect the time-axis because of the same inertia,
or our description in terms of a field $\epsilon$ breaks down before the intersection  takes place. }.

Another way to obtain reheating (without invoking two field dynamics) is to postulate a phase transition once $\Phi$ has reached some sufficiently large value, and to assume that at that point the energy in $\Phi$ is released as thermal energy. It may also
be possible to implement the ``warm inflation" scenario \cite{Berera} in which the dynamics of $\Phi$ during inflation is driven by a thermal friction term $\Gamma {\dot \Phi}$ in eq. (\ref{9}) which dominates over the Hubble damping term, in which case there is continuous heating during inflation. However, we feel that the two field model described above is the most promising.

We believe that other problems which are
inherent features of most inflationary models
can also find their  natural solutions in the approach suggested
in this Letter. In particular,
let us briefly mention the ``fluctuation problem''.
The only known solutions of this problem in the context of scalar field driven inflationary models involve introducing very small parameters
into the scalar field potential. Otherwise, the
predicted quantum fluctuations are in excess of those
allowed by the bounds on cosmic microwave background (CMB) anisotropies.

If we were to treat $\Phi$ as a fundamental scalar field and to apply the usual theory of quantum generation and classical evolution of fluctuations \cite{Review}, \cite{Flucts}, we would also encounter the ``fluctuation problem" in our scenario. The magnitude $\delta T / T$
of the predicted fluctuations in the CMB would be given by
\be
\label{18}
{{\delta T} \over T} \sim {{H V^{\prime}} \over {{\dot \Phi}^2}},
\ee
where the quantities are evaluated during the inflationary epoch at the time when the length scales which correspond to the angular scales of the CMB anisotropies being considered leave the Hubble radius. During the era of
``slow rolling", we can make use of the equation of motion (\ref{9}) with
${\ddot \Phi} = 0$ and equations (\ref{10}) and (\ref{12}) to estimate the quantities appearing in (\ref{18}). Introducing the dimensionless constant $c$ via ${\tilde c} = c m_{pl}^5$ we obtain the following estimate:
\be
\label{19}
{{\delta T} \over T} \sim 10^2 f(1)^{1/2} c^{1/2}.
\ee
Thus, unless $f(1)$ happens to be extremely small (there is no reason to expect this to be the case), Planck scale inflation ($c \sim 1$) seems to lead to the usual fluctuation problem in that $\delta T / T$ is greater than one instead of $10^{-5}$.
The reason for the fluctuations to be large is clear: at the Planck scale
we have only one dimensional parameter.
This parameter governs
the dynamics of fluctuations. With the parameter of order one (in Planck units), an amplitude of fluctuations of order one will result.

In QCD we would encounter the same problem if we were to start from the wrong
vacuum state. However, in QCD we know how to handle this problem, at least qualitatively\footnote{If the dynamics of $\Phi$ were dominated by friction instead of by the Hubble expansion, then the ``fluctuation problem" could also be resolved as in Ref. \cite{Berera} since the dominant fluctuations are thermal and not quantum.}
We separate the strong and weak fluctuations. The strong fluctuations,
which are presumably responsible  for all nonperturbative phenomena,
can be taken into account phenomenologically: they form a nontrivial background
and nontrivial condensates which can not be calculated at the moment, but
should be extracted from  experiment. The weak fluctuations
(which is the only relevant issue at the moment)
in this background are in fact weak: standard perturbative techniques
and the Wilson Operator Expansion
are appropriate methods to calculate them.

The moral is simple: if one manages to describe the
main vacuum background fields correctly, the quantum fluctuations
in this true background will presumably be small, and it is
these vacuum fluctuations
in the gravitational field which will
generate the CMB anisotropies.
A strong quasi-classical vacuum configuration at the Planck scale
is clearly of order of one, however this  is  not the relevant configuration
for the analysis of the CMB fluctuations.
Of course, it is difficult to check such a treatment of the
quantum fluctuations in gravity
because we do not understand the quantum dynamics of the theory.

\section{Conclusion}

We have proposed an inflationary scenario in which inflation is driven by
the dynamics of a condensate of gravitational fields which forms at some
very early time in the evolution of the Universe (presumably at the Planck
scale). We have interpreted the vacuum energy of this condensate as the effective potential which determines the dynamics of the condensate. The
effective potential in turn is calculated using an asymptotic series
approach discussed previously in Ref. \cite{ARZ}.

We have shown that the ``slow rolling" conditions (a key requirement for
successful inflation) are automatically satisfied. The same dynamics of
the infrared cutoff which is responsible for relaxing any contribution to the cosmological constant due to this condensate leads to a termination of the
period of inflation and thus to a graceful exit from inflation
\footnote{In this
respect, there are analogies with the proposal of Ref. \cite{Woodard}.
However there are also differences: in  Ref. \cite{Woodard}
the infrared divergences act uniformly in space, whereas
in our case the infrared dynamics is represented
by a function $\epsilon (t)$ which could in principle also depend on space.  
In general, this function need not be small.
However,
if inflation is successful, then the same
inertia responsible for this success drives the function $\epsilon$
(and with it also the cosmological constant)
to zero. A different behavior of
$\epsilon (t)$ might lead to an empty and cold Universe with a nonzero
cosmological constant.}.
We have also
made some speculations concerning a mechanism by which the ``fluctuation
problem" could be solved. This mechanism relies on the fact that the
field which drives inflation in our scenario is not a fundamental field, but rather a condensate with a complicated vacuum structure (like in QCD).

One of the nice features of our proposal is that inflation is tied to the
way in which (at least part of) the cosmological constant problem is
resolved. We believe that this is the most important new ingredient
in the proposed inflationary scenario.
\bigskip

\centerline{\bf Acknowledgments}

For useful discussions we are grateful to Andrei Linde,
Paul Mende and Richard Woodard. One of us (R.B.)
wishes to thank the Physics Department of the University of British Columbia
for its hospitality during a visit when this work was initiated. The work of
R.B. is supported in part by the U.S. Department of Energy under Grant
DE-FG0291ER40688, Task A, the work of A.Z. is supported in part
by the Natural Sciences and Engineering Research Council of Canada.

\end{document}